\documentclass[%
 reprint,
 amsmath,amssymb,
 aps,prd,
floatfix,
]{revtex4-2}

\usepackage{graphicx}
\usepackage{dcolumn}
\usepackage{bm}
\usepackage{xcolor}
\usepackage{float}

\usepackage{hyperref}

\begin{document}



\title{Topological features of the deconfinement transition}

\author{S. Bors\'anyi} \email{borsanyi@uni-wuppertal.de}
\affiliation{University of Wuppertal, Department of Physics, Wuppertal D-42119, Germany}

\author{Z. Fodor}
\affiliation{Pennsylvania State University, Department of Physics, University Park, PA 16802, USA}
\affiliation{University of Wuppertal, Department of Physics, Wuppertal D-42119, Germany}
\affiliation{Inst.  for Theoretical Physics, ELTE E\"otv\"os Lor\' and University, P\'azm\'any P. s\'et\'any 1/A, H-1117 Budapest, Hungary}
\affiliation{J\"ulich Supercomputing Centre, Forschungszentrum J\"ulich, D-52425 J\"ulich, Germany}

\author{D. A. Godzieba}
\affiliation{Pennsylvania State University, Department of Physics, University Park, PA 16802, USA}

\author{R. Kara}
\affiliation{University of Wuppertal, Department of Physics, Wuppertal D-42119, Germany}

\author{P. Parotto}
\affiliation{Pennsylvania State University, Department of Physics, University Park, PA 16802, USA}

\author{R. Vig} \email{tajhajlito@gmail.com}
\affiliation{University of Wuppertal, Department of Physics, Wuppertal D-42119, Germany}

\author{D. Sexty}
\affiliation{University of Graz, NAWI Graz, Institute for Physics, A-8010 Graz, Austria}

\date{\today}


\begin{abstract}
The first order transition between the confining and the center symmetry breaking
phases of the SU(3) Yang-Mills theory is marked by discontinuities in
various thermodynamics functions, such as the energy density or the
value of the Polyakov loop. We investigate the non-analytical behaviour 
of the topological susceptibility and its higher cumulant around
the transition temperature and make the connection to the curvature
of the phase diagram in the $T-\theta$ plane and to the latent heat.
\end{abstract}

\maketitle



\section{\label{sec:Introduction}Introduction} 

Quantum Chromodynamics, the theory of strong interactions, features
a broad cross-over around 155 MeV temperature \cite{Aoki:2006we,Bazavov:2018mes,Borsanyi:2020fev}. 
In the high temperature phase chiral symmetry is restored,
and quarks are no longer localized in hadrons. The order of this 
deconfining transition depends on the values of the quark masses
\cite{Pisarski:1983ms,Brown:1990ev,Cuteri:2020yke,Cuteri:2021ikv}.
 The best studied special case
is the theory with infinite quark masses, where a first order
transition was predicted by renormalization group arguments
\cite{Svetitsky:1982gs,Yaffe:1982qf} and later shown numerically on the lattice
\cite{Brown:1988qe,Fukugita:1989yb,Borsanyi:2022xml}.

The topological features of hot QCD matter came into focus mainly
because of their impact on axion search experiments
\cite{Irastorza:2018dyq}. Axions are hypothetical particles linked to
the Peccei-Quinn mechanism, a proposed solution to the strong CP
problem \cite{Weinberg:1977ma,Wilczek:1977pj,Peccei:1977hh}, which are also candidate dark matter constituents.
Their abundance in our present world is determined by the temperature
of the hot early Universe at the point of their production: 
lighter axions are produced at a later
stage in a colder Universe, resulting in a larger density today, because of the shorter period of expansion in comparison to a heavier axion that would have had more time for dilution~\cite{Wantz:2009it}. 
It must be noted that this picture is not complete without the details of the production mechanism,
such as the interplay with the formation and decay of global cosmic strings
\cite{Klaer:2017ond,Buschmann:2021sdq}.

Lattice QCD has provided essential input for constraining a class of axions,
the QCD axion. In particular, the relation between the temperature and the
axion mass was determined up to a constant factor in a broad temperature
range \cite{Petreczky:2016vrs,Borsanyi:2016ksw}. The mass of the QCD
axion is controlled by the strongly temperature dependent topological
fluctuations. The relevant observable is the topological susceptibility
\begin{equation}
\chi = \frac{\langle Q^2\rangle}{\mathcal{V}} \label{eqn:chi}
\end{equation}
where $\mathcal{V}$ is the Euclidean four-volume and $Q$ is the topological
charge, defined in the continuum theory as
\begin{equation}
Q = \frac{1}{32\pi^2} \int_{\mathcal{V}} d^4x \mathrm{Tr}\, \epsilon_{\mu\nu\rho\sigma} 
F_{\mu\nu} F_{\rho\sigma} \, \, .
\label{eq:FFtilde}
\end{equation}
Analytic arguments require a power-law drop of $\chi$ 
with logarithmic corrections as the temperature is increased 
in the weak coupling regime \cite{Gross:1980br}. This behaviour was, indeed,
observed in exploratory studies on the lattice
\cite{Hoek:1986nd,Teper:1988ed} both in the quenched case \cite{Alles:1996nm}
and in full QCD \cite{Alles:2000cg}. While most studies used the
field theoretic definition of $Q$ of Eq.~(\ref{eq:FFtilde}), the 
conclusion was unaltered by using the index theorem to define $Q$
\cite{Gattringer:2002mr}.

The precise value of the susceptibility in the high temperature phase
determines the axion potential in the hot early Universe. Its calculation at
high temperatures is challenging even in the quarkless SU(3) theory
and required large scale studies \cite{Borsanyi:2015cka}. 
Research within the lattice QCD community was pursued in several directions:
i) to calculate the susceptibility at high temperatures, ii) to address higher
cumulants, and iii) to understand the topological features in the context
of the large-$N$ limit of the SU(N) theory.

The smallness of $\chi(T)$ at the axion production temperature (which can be several GeVs)
requires the computation of $\langle Q^2\rangle$ based on extremely rare
events. This means, that the variance of the integer valued charge $Q$ has
to be determined, which can be several orders of magnitude below one, and most of the sampling will result in $Q=0$ on the course of a simulation.
The integral method was first suggested to mitigate this algorithmic challenge
\cite{Frison:2016vuc,Borsanyi:2016ksw}.
It extracts $\chi$ from the difference of the free energy between the
$Q=0$ and $Q=1$ sectors. Other
ideas include metadynamics \cite{Jahn:2018dke,Jahn:2020oqf},
a multicanonical approach \cite{Bonati:2018blm},
and density of states methods \cite{Gattringer:2020mbf,Borsanyi:2021gqg}.
In the case of dynamical QCD the calculation of the susceptibility
was further refined by using the spectral features of the Dirac 
operator \cite{Athenodorou:2022aay}, or by using reweighting and eliminating known
staggered artefacts \cite{Borsanyi:2016ksw} to reduce cut-off effects.

In the temperature region around and below the transition, the computation of
higher moments of the topological charge requires very high statistics
\cite{Bonati:2013tt}. For instance, the kurtosis is expressed through
the $b_2$ coefficient:
\begin{equation}
b_2(T)= -\frac{\chi_4(T)}{12\chi(T)}\,,\qquad
\chi_4=\frac{1}{\mathcal{V}}
\left[ \left\langle Q^4\right\rangle - 3 \left\langle Q^2\right\rangle^2\right] \, \, .
\label{eq:b2}
\end{equation}
It was observed, that simulations at imaginary values of
a $\theta$ parameter are feasible without the emergence of a sign problem,
and enable the precise study of higher cumulants of the topological
charge at $\theta=0$ \cite{Panagopoulos:2011rb,Bonati:2015sqt}.

The SU(3) theory can be seen as a special case of the SU(N) gauge theories,
and it can be studied in the framework of the large-$N$ expansion
\cite{Vicari:2008jw}. In the large-$N$ limit the topological susceptibility
is constant up to the deconfinement temperature, but on the high temperature
side of the transition it is suppressed exponentially with $N$
\cite{Lucini:2004yh,DelDebbio:2004vxo}, in agreement with the semiclassical
expectations \cite{Kharzeev:1998kz}.

The topological susceptibility $\chi$ is the second derivative of the thermodynamic
potential with respect to the $CP$-breaking $\theta$ parameter. $\chi$
and the higher moments are the Taylor coefficients of the QCD
pressure when it is extrapolated to non-zero $\theta$. It is, thus, to
be expected, that the behaviour of the susceptibility near the transition
is linked to the details of the phase diagram in the $T - \theta$ plane.
It was pointed out in Refs.~\cite{DElia:2012pvq,DElia:2013eua} that in the
case of a first order transition the curvature parameter $R_\theta$, defined as
\begin{equation}
\frac{T_c(\theta)}{T_c(0)}=1+ R_\theta \theta^2 + \mathcal{O}(\theta^4)
\label{eq:Rtheta}
\end{equation}
is related to the latent heat $\Delta\epsilon$ and the discontinuity of
the topological susceptibility across the transition $\Delta\chi$ by
a Clausius–Clapeyron--like equation
\begin{equation}
\Delta\chi = 2 \Delta\epsilon R_\theta \, \, .
\label{eq:clapeyron}
\end{equation}

To make our discussion self-contained we revisit its derivation here. The first order deconfinement transition 
can be described with the free energy densities of the two phases of the system, $f_c(T)$ and $f_d(T)$, which are equal at the transition point,
and the difference of their temperature derivatives is connected to the latent heat $\Delta \epsilon$
\begin{equation}
 \Delta \epsilon = \epsilon_d-\epsilon_c = \left. T^2 \left( -\partial_T( f_d/T) +\partial_T ( f_c/T) \right) \right|_{T=T_c}  
\end{equation}
Near the transition, the free energy densities (using the reduced temperature $(t)$) are approximated as 
\begin{equation}
    f_{\alpha}(t,\theta) = f_0 + T_{c} A_\alpha t + { \chi_\alpha \over 2 } \theta^2
\end{equation}
where we have neglected 
higher order terms,
and $ \Delta \epsilon= T_c (A_c-A_d)$.
At finite $\theta$, the coincidence of $f_c(t,\theta)$ and $f_d(t,\theta)$ signifies the shifted transition temperature,
yielding the equation
\begin{equation}
 A_c t +  { \chi_c \over 2 T_c} \theta^2 = A_d t + { \chi_d \over 2 T_c } \theta^2 
\end{equation}
which simplifies to
\begin{equation}
   {T_c(\theta) \over T_c(0) } = 1 + { \chi_d - \chi_c \over 2 \Delta \epsilon }  \theta^2  
\end{equation}
proving the relation $ R_\theta = \Delta \chi / 2 \Delta \epsilon$.

If the susceptibility drops in value as the transition is traversed from the
cold, confined phase, the curvature $R_\theta$ must be negative.
$R_\theta$ was extracted from the dependence of the transition temperature
at various imaginary $\theta$ values in a large scale lattice study
which yielded $R_\theta=-0.0178(5)$ \cite{DElia:2012pvq,DElia:2013eua}.

In a recent work we used an algorithmic development,
parallel tempering, to reach higher precision of the latent heat
of the SU(3) Yang-Mills theory \cite{Borsanyi:2022xml}. Our result 
$\Delta\epsilon/T_c^4 = 1.025(21)(27)$ can be combined with
\cite{DElia:2013eua} to find $\Delta\chi/T_c^4 =- 0.0365(18)$,
which corresponds to an error of 5\%.

The goal of this work is to quantify the discontinuity
$\Delta\chi$ as a direct lattice result. The topological features near $T_c$
were rarely addressed in the continuum limit in the existing
literature, and finite volume scaling was often neglected.
After introducing the lattice setup in section \ref{sec:Q}, we
show high-statistics results for the basic observables, $\chi(T)$
and $b_2(T)$ in section \ref{sec:stopo}. In sections \ref{sec:jump}
and \ref{sec:curvature} we calculate the continuum and infinite volume
limits of $\Delta\chi$ and $R_\theta$, respectively.
In the discussion of section \ref{sec:b2} we give
an account of the fourth moment near $T_c$ to complement earlier works that report an early onset of the dilute instanton gas picture \cite{Vig:2021oyt}.


\section{\label{sec:Q}The topological charge on the lattice}

According to Eq.~\eqref{eqn:chi}, in order to directly obtain $\Delta\chi$ we needed to determine the lattice version of $Q$ corresponding to gauge configurations of different ensembles generated at the transition point. We simulated the pure SU(3) Yang-Mills theory with the Symanzik-improved gauge action in narrow range of gauge couplings around $\beta_c$ using parallel tempering. The center of this range was fine tuned to the critical coupling $\beta_c$ with a per mille precision in $T/T_c$, at this coupling we stored the configurations for further analysis. The use of parallel tempering has significantly reduced the auto-correlation time by allowing a frequent exchange of configurations between $T<T_c$, $T\approx T_c$ and $T>T_c$ sub-ensembles.
The number of gauge configurations stored and later evaluated at $\beta_c$ are summarized in table \ref{tab:1}.
\begin{table}[ht]
\centering
\begin{tabular}{rl|l|l|l|l|l|}
\cline{3-7}
&&\multicolumn{5}{c|}{$N_\tau$}\\\cline{3-7} \vspace{-9pt} \\ \cline{3-7}
&&  6& 7& 8 & 10&12  \\ \cline{1-7}
\multicolumn{1}{|r}{\vline \thinspace \vline} &$2$& 18977  & 13055& 10098  & 8552 &11882 \\
\multicolumn{1}{|r}{\vline \thinspace \vline} &$4$& 
49747 
& 64901 & 77902 & 40054&20604 \\
\multicolumn{1}{|r}{\vline \thinspace \vline}  &$4.5$& -  & - & 30544 & -& -\\
\multicolumn{1}{|r}{  $LT$ \vline \thinspace \vline}&$5$&  20041&  6524& 36610 & 13473&- \\
\multicolumn{1}{|r}{\vline \thinspace \vline}  &$6$&  67185&  7875& 53325 & 24475&- \\
\multicolumn{1}{|r}{\vline \thinspace \vline}  &$8$&  30581&  6677& 7372 &- & -\\ \hline
\end{tabular}
\caption{\label{tab:1}Number of gauge configurations generated at the transition point. $LT=N_x/N_\tau$ means the aspect ratio and $N_x$ and $N_\tau$ are the spatial and the temporal extensions in lattice units.}
\end{table}

\begin{figure}
    \centering
    \includegraphics[width=0.4\textwidth]{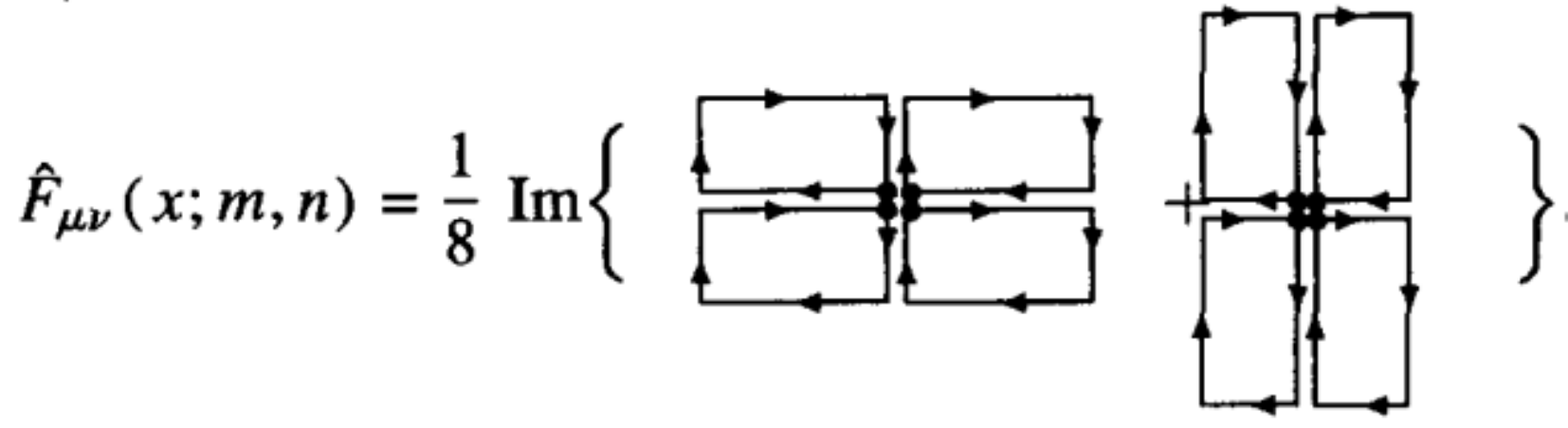}
    \caption{$1\times 2 $ plaquettes in the improved clover discretisation of the topological charge.}
    \label{fig:F_mu_nu}
\end{figure}

On each lattice configuration we measured the Symanzik-improved topological charge defined similarly as in  \cite{Moore:1996wn,deForcrand:1997esx,BilsonThompson:2002jk}
\begin{equation}
Q=\sum_{mn \in \{11,12\}} c_{nm} Q_{mn} 
\end{equation}
where the coefficients $c_{mn}$ are 
\begin{equation*}
c_{11}=10/3, \;\;\; c_{12}=-1/3
\end{equation*}
and $Q_{mn}$ is the naive topological charge defined through the lattice version of the field strength tensor ($\hat F_{\mu\nu}$)
\begin{multline}
Q_{mn}=\frac{1}{32\pi^2}\frac{1}{m^2n^2}\sum_x \sum_{\mu,\nu,\rho,\sigma} \epsilon_{\mu\nu\rho\sigma} \cdot \\ \cdot {\rm Tr}(\hat{F}_{\mu\nu}(x;m,n)\hat{F}_{\rho\sigma}(x;m,n)).
\end{multline}
$\hat F_{\mu\nu}(x;m,n)$ is built by averaging clover terms of $m\times n$ plaquettes at site $x$ on the $\mu\nu$ plane. We use Figure \ref{fig:F_mu_nu} from \cite{deForcrand:1997esx} as a visualization of $\hat F_{\mu\nu}(x;m,n)$. 

We introduced smearing on the gauge field via the Wilson flow, which allowed us to measure a renormalized topological charge which we defined at a given flow time $t$. We intergrated the Wilson flow using a 3rd order adaptive step-size variant of the Runge-Kutta scheme in Ref.~\cite{Luscher:2010iy}.
All moments of $Q$ are a constant function of the flow time $t$ in the continuum. In practice one selects a fixed flow time $t$ in physical units, e.g. relative to the actual temperature $T$ at which the continuum extrapolation can be carried out using the lattices at hand. The choice of $t$ is, thus, a compromise, such that $t$ should be small enough to avoid high computational costs but also to avoid finite volume effects $t\ll L^2$, yet large enough to maintain $t\gg a^2$.

\begin{figure}
    \centering
    \includegraphics[width=0.48\textwidth]{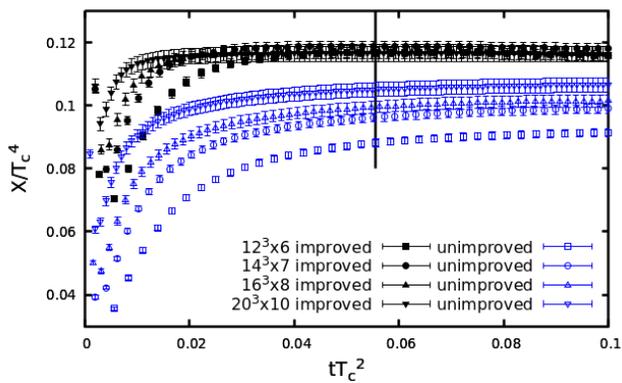}
    \caption{Topological susceptibility calculated on lattices of aspect ratio $LT=2$ with different resolutions. Data represented with black filled points  are determined from the Symanzik-improved topological charge compared to unimproved data shown as blue empty points. }
    \label{fig:chi-t}
\end{figure}

\begin{figure}
    \centering
     \includegraphics[width=0.48\textwidth]{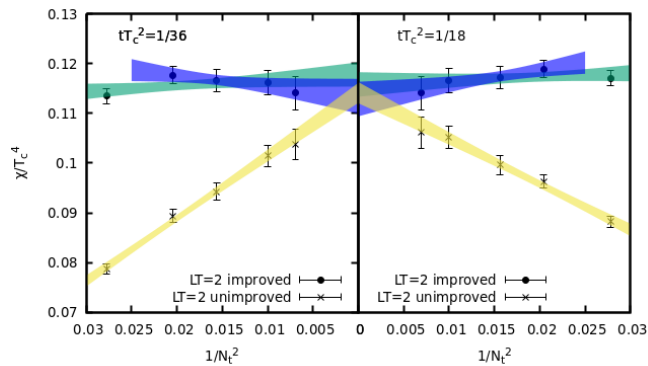}
    \caption{Continuum extrapolations of the topological susceptibility at the transition temperature calculated on lattices with aspect ratio $LT=2$. On the right hand side we show results calculated from improved and unimproved charge defined at $t=1/18$. On the left hand side we show the data obtained in the case of $t=1/36$. The colored bands show linear fits on the data. The shorter bands of color blue are linear fits on the improved data using data only from finer lattices $N_\tau>6$. In the case of $t/T_c^2$=1/36 the reduced chi square $\chi^2_r=\chi^2$/(degrees of freedom) of the fits for the unimproved, improved and the short range improved data are respectively $\chi^2_{\texttt{unimp}}=1.06/3$, $\chi^2_{\texttt{imp}}=3.25/3$ and $\chi^2_{\texttt{imp,s}}=0.14/2$. In the case of $t/T_c^2$=1/18 we got $\chi^2_{\texttt{unimp}}=0.85/3$, $\chi^2_{\texttt{imp}}=1.57/3$ and $\chi^2_{\texttt{imp,s}}=0.17/2$.}
    \label{fig:chi-Nt}
\end{figure}

To make a practical choice for this study we examined the $t$ dependence of $\chi$, which can be seen in Fig.~\ref{fig:chi-t}. In the figures we show the normalized susceptibility $\chi/T_c^4=(T/T_c)^4\left<Q\right>(N_\tau^4/\mathcal{V}) $, so that the comparison of lattices with different resolution is meaningful. Different curves within the same color represent different lattice spacings, and with different colors we show data that were calculated from the improved or unimproved topological charge. We defined $Q$ at a flow time that fell into the plateau region even in the case of the coarsest lattices. Our choice of the flow time $tT^2=1/18$ is highlighted with a black vertical line in Fig.~\ref{fig:chi-t}. Fixing $t$ we could calculate $\chi$ and determine a continuum limit for Symanzik-improved and unimproved data sets, which we compare in Fig.~\ref{fig:chi-Nt}, together with results that we calculated at a smaller flow time $tT^2=1/36$. With $Q$ that is renormalized correctly the continuum limits obtained from improved and unimproved data should agree. This is true in the case with our choice of $t$ (right hand side), whereas at a smaller $t$ (left hand side) finite size effects are still significant and improved and unimproved continuum extrapolations slightly differ. The blue bands show a shorter range fit on the improved data, excluding the $N_\tau=6$ lattice. The continuum limit from the smaller fit range extrapolation is compatible with results using the whole data set, therefore we will use one or the other of the two cases in the following sections depending on the $\chi^2$ of the fit on the actual data. 


In the next section we describe a broader temperature scan throughout the transition $0.9 ~ T/T_c$-$1.1 ~ T/T_c$. Calculating the Wilson flow at several temperatures would have a large computational cost, therefore in this case we calculated $Q$ after using stout smearing on the gauge field corresponding to the same physical smearing radius as in the case of the Wilson flow. 

In practice, we performed a number of stout smearings ($\rho=0.125$), such that $t/T_c^2=1/18 = N_{\text{smear}\rho}/N_\tau^2$. This means
$N_{\text{smear}}=16$ for the coarsest lattice ($N_\tau=6$) and $21.\bar{7}$ steps for $N_\tau=7$, $28.\bar{4}$ for $N_\tau=8$, and $44.\bar{4}$ steps for $N_\tau=10$. Non integers steps were realized through
an interpolation of $Q$ in the step number.

To determine the systematic error coming from using an alternative cooling method we calculated $\chi$ both ways from the $LT=2$ lattice data. Results are shown in table \ref{tab:2}. There is a precise agreement in $\chi$ calculated in the two different cases at all $N_\tau$ values. This justifies the use of stout smearing on configurations generated at several temperatures. We also used stout smearing on configurations of imaginary $\theta$ simulations discussed in sections \ref{sec:curvature} and \ref{sec:b2}. 

\begin{table}[ht]
\centering
\begin{tabular}{|l|l|l|}
\hline
&\multicolumn{2}{c|}{$\chi/T_c^4$}\\
\cline{2-3}
\multicolumn{1}{|r|}{$N_\tau$}&Wilson flow&stout smearing\\ \hline
\multicolumn{1}{|r|}{6}&0.11702(156)&0.11718(155)\\
\multicolumn{1}{|r|}{7}&0.11882(176)&0.11884(176)\\
\multicolumn{1}{|r|}{8}&0.11720(231)&0.11722(233)\\
\multicolumn{1}{|r|}{10}&0.11652(248)&0.11655(248)\\
\multicolumn{1}{|r|}{12}&0.11416(311)&0.11413(311)\\\hline
\end{tabular}
\caption{\label{tab:2} Topological susceptibility calculated at $t/T_c^2=1/18$ at temporal extents $N_\tau=6,7,8,10,12$ via the Wilson flow (second column) compared to $\chi$ calculated after stout smearing steps (third column) corresponding to the same physical flow time.}
\end{table}

\section{\label{sec:stopo}The susceptibility and $b_2(T)$ in the transition region}

In addition to the simulations we carried out by tuning precisely at the transition temperature, we measured $Q$ for ensembles generated in the vicinity of the transition temperature. Employing parallel tempering -- as in our recent work \cite{Borsanyi:2022xml} -- we were able to cover the temperature range $0.9 T_c < T < 1.1 T_c$ in a fine mesh of 64 or more gauge couplings.
In the previous section we observed that the topological susceptibility can be extracted both from the flow based definition and through a sequence of stout smearings ($\rho=0.125$), the difference between the two methods is statistically insignificant. 
We perform a temperature scan, evaluating $Q$ at 64 or more temperatures, thus, we opted for the cheaper smearing sequence.

Obtaining the charge $Q$ allowed us to examine the temperature dependence of both the topological susceptibility and $b_2$ in the transition region. We report in this Section our results for the aspect ratio $LT=4$, in which case we could carry out the continuum extrapolation that we show in the following.

\begin{figure}
    \centering
    \includegraphics[width=0.5\textwidth]{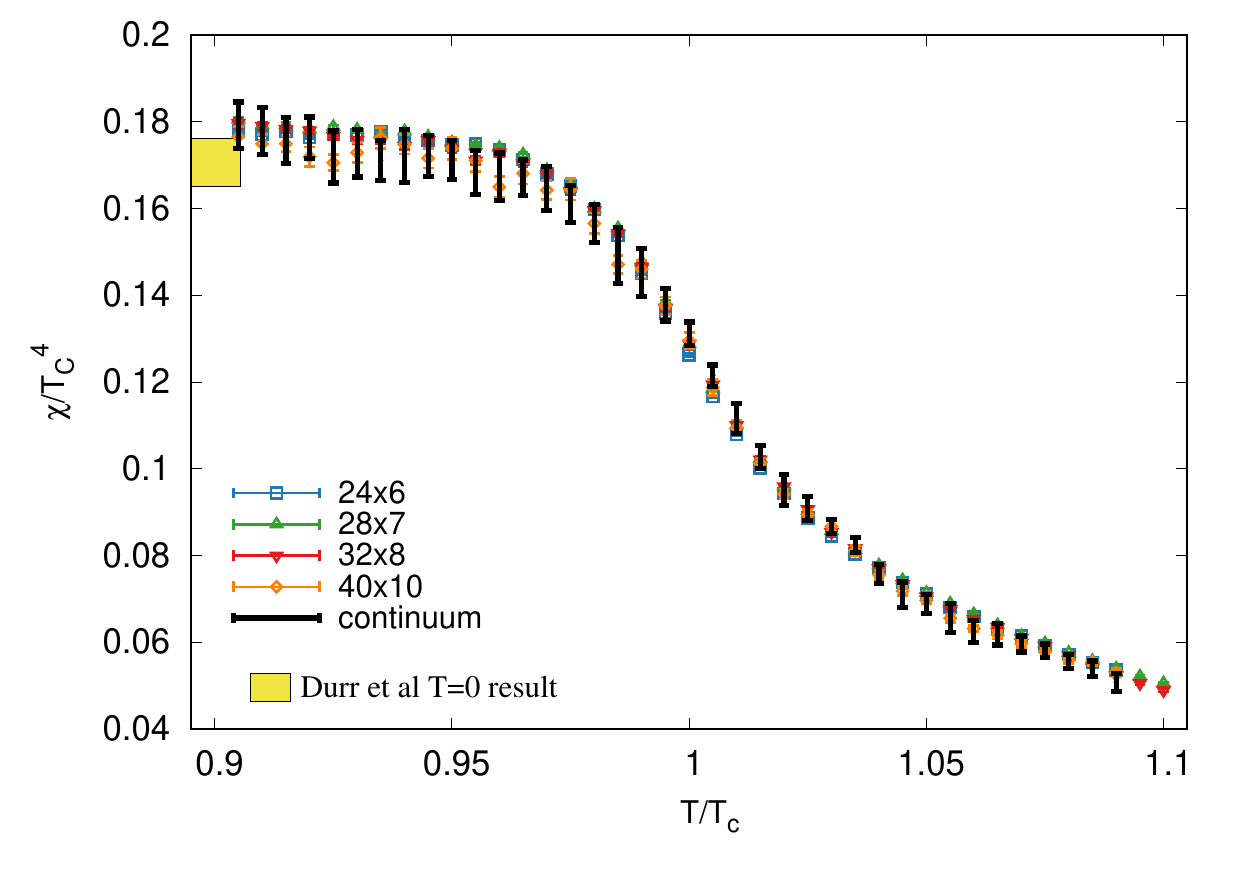}
    \includegraphics[width=0.5\textwidth]{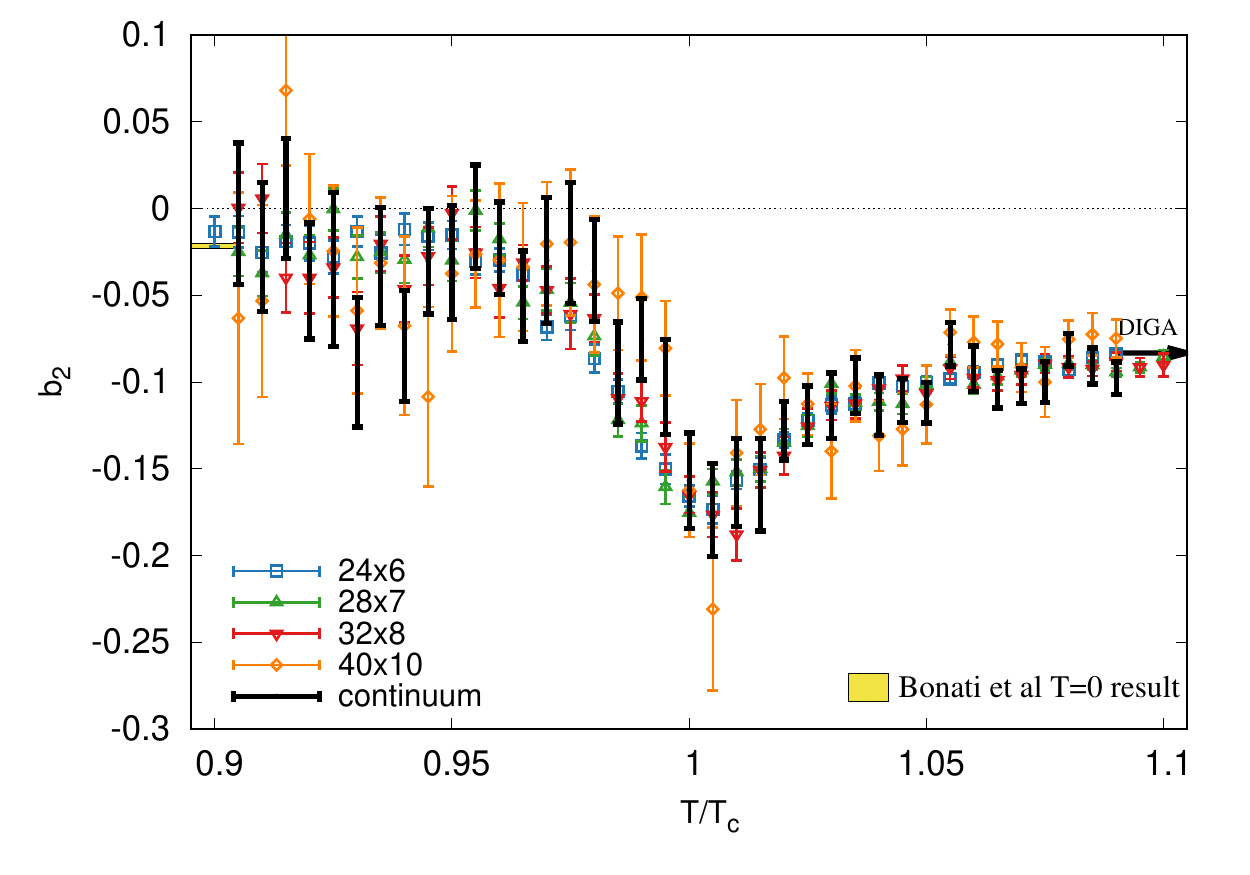}
    \caption{Normalized topological susceptibility (top) and $b_2$ coefficient as functions of the normalized temperature. Results for lattices with physical volume $LT=4$ and $N_\tau=6,7,8,10$ are shown in blue, green, red and orange respectively. The continuum extrapolation, which includes statistical and systematic uncertainties, is shown in black.
    For each quantity we quote a zero temperature result from the literature, the precision study for $\chi$ from \cite{Durr:2006ky} and the imaginary-$\theta$ based result of \cite{Bonati:2015sqt} for $b_2$.
    }
    \label{fig:chiANDb2_LT4_cont}
\end{figure}

In Fig.~\ref{fig:chiANDb2_LT4_cont} we show the normalized topological susceptibility $\chi/T_c^4$ (top panel) and the coefficient $b_2$ (bottom panel). The colored points correspond to lattices with $LT=4$ and $N_\tau=6,7,8,10$. In order to carry out a continuum extrapolation, we used a spline interpolation in the gauge coupling to extract data at equal temperatures for all $N_\tau$. The gauge couplings we actually used
depend on the scale setting choice. We analyzed our
data with two different scale setting functions ($T_ca(\beta)$). For both settings we relied on the results of our previous project \cite{Borsanyi:2022xml}.

The first scale setting is defined through $w_0$:
we used the $w_0/a(\beta)$ data set from $48^4$ lattice simulations in the same $\beta$ range. These $w_0/a$ data were translated to $T_ca(\beta)$ scale by the factor  $w_0T_c=0.25265$ valid for the aspect ratio $LT=4$ (and neglecting its per-mill level error).

The second scale setting was defined through
the sequence
of the transition gauge couplings ($\beta_c(N_\tau)$) for various $N_\tau$ as determined in Ref.~\cite{Borsanyi:2022xml}. We, thus, set $T_ca(\beta_c(N_\tau))=1/N_\tau$ and interpolate
to the other gauge couplings (using a polynomial fit).

The continuum limit is then performed at fixed
temperatures independently. For the susceptibility
the statistical errors are small enough on all lattices,
and we can estimate the systematic error of the continuum extrapolation by first fitting with the $N_\tau=6$
and omitting it in a second fit. The statistical errors
on the $b_2(T)$ result on the finest lattice ($40^3\times10$)
is too large for this estimation for $b_2$; there
all four lattices were included in the continuum limit.

In Fig.~\ref{fig:chiANDb2_LT4_cont} we also show
the corresponding $T=0$ results. Durr et al. presented their result in $r_0$ units \cite{Durr:2006ky}.
We combined $w_0T_c=0.25384(23)$ from our recent Ref.~\cite{Borsanyi:2022xml} with $w_0/r_0=0.341(2)$
from Ref.~\cite{Sommer:2014mea} and obtained $\chi/T_c^4=0.1707(55)$. This is in agreement with
newer continuum results of Athenodorou\&Teper $\chi/T_c^4=0.18(1)$ \cite{Athenodorou:2020ani} and Bonati et al. $\chi/T_c^4=0.16(1)$ \cite{Bonati:2015sqt}. The $b_2(T=0)=-0.0216(15)$ continuum result is from Ref.~\cite{Bonati:2015sqt}.


\section{\label{sec:jump}The discontinuity of the topological susceptibility}

The rapid drop of $\chi(T)$ near $T_c$ is well known from early lattice works \cite{Alles:1996nm}. The strong temperature dependence is a characteristic feature throughout the high temperature phase. To quantify the discontinuity ($\Delta\chi(T_c)$) at $T_c$ one requires a dedicated study complete with continuum limit and volume extrapolation. This is the subject of the present section.

\begin{figure}[h]
    \centering
    \includegraphics[width=0.5\textwidth]{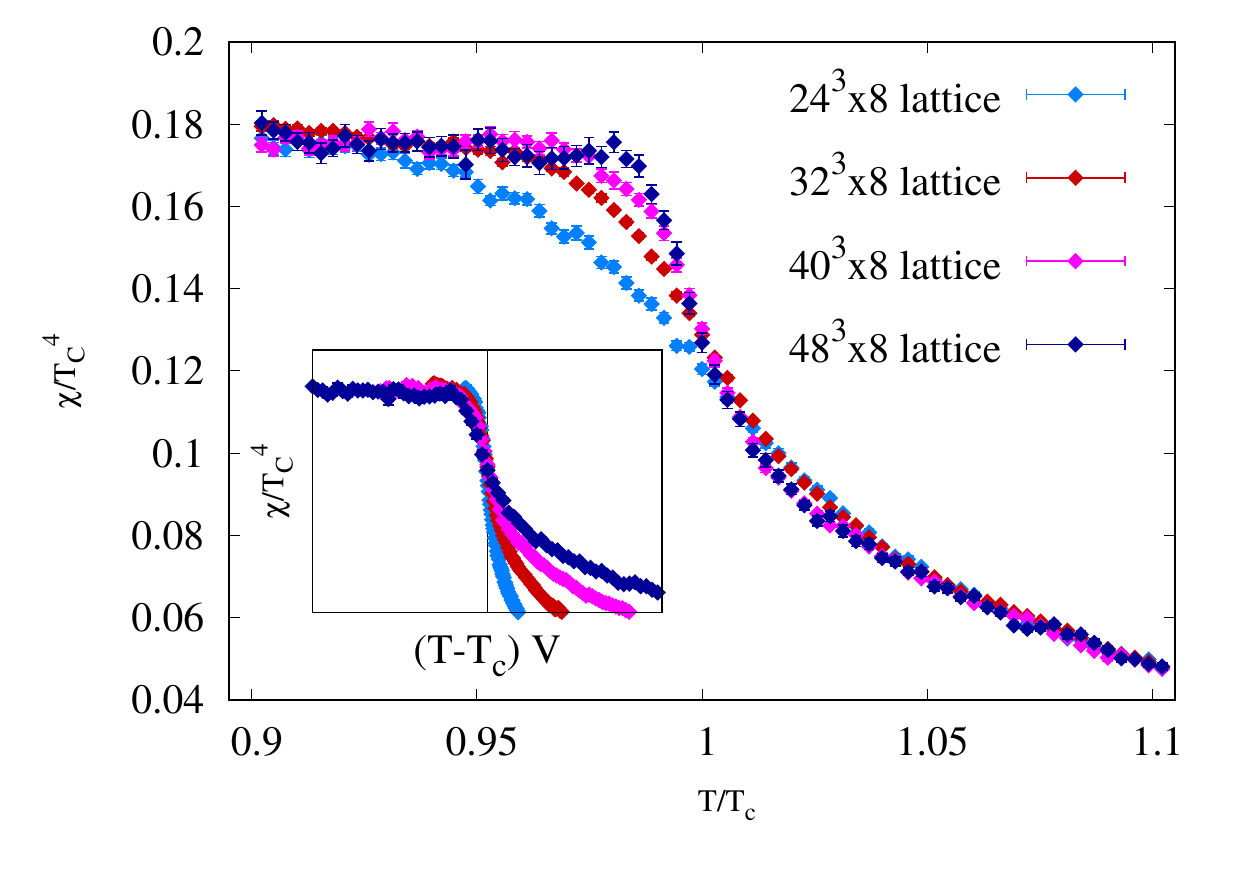}
    \includegraphics[width=0.5\textwidth]{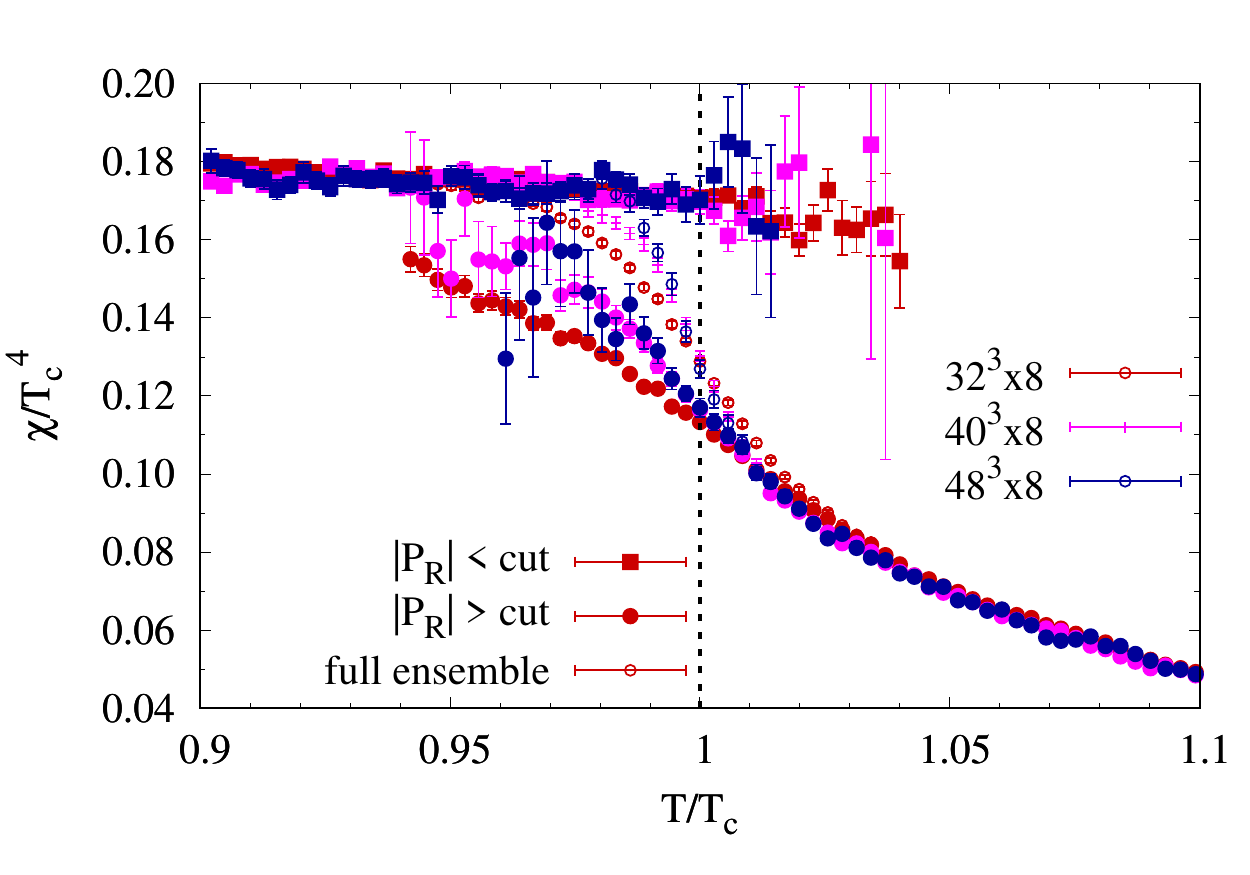}
    \caption{Normalized topological susceptibility
    as a function of temperature near $T_c$. In the top panel we compare four volumes. The increasing slope indicates a discontinuity. The inset plot normalizes the temperature axis with the volume: there the curves overlap at and below $T_c$.
    The bottom panel shows the susceptibility
    for the same runs, but the high and low temperature
    phases were separated into two sub-ensembles.
    }
    \label{fig:chicloseup}
\end{figure}

We start with showing lattice data for $\chi(T)/T_c^4$
using four different aspect ratios $LT=3,4,5$ and 6
for one $N_\tau$ in Fig~\ref{fig:chicloseup}. 
The curves behave visibly differently
below and above $T_c$. In the deconfined phase we see
no significant volume dependence, but below $T_c$ the slope rapidly grows with the volume. The inset plot shows this on a rescaled temperature axis. The approximate overlap of the $\chi/T_c^4$ curves then
is a manifestation of the discontinuity at the
temperature of the first order transition.

For the lower panel of Fig.~\ref{fig:chicloseup} we analyzed the same lattice configurations by splitting the ensembles into confined ($|P|<P_c$) and deconfined ($|P|>P_c$) sub-ensembles, where $P_c$ is a suitable cut in the Polyakov loop absolute value. This splitting is an ambiguous procedure away from $T_c$ and for finite volumes. For simplicity we let $P_c$ be the position of the local minimum
of the renormalized Polyakov loop histogram at $T_c$
for all temperatures. We see that, at $T_c$, $\chi$ takes very distinct values in the two phases, and this extends to a small vicinity of the transition temperature, depending on the volume.

The splitting method has already been used several times
in the literature to calculate the latent heat \cite{Shirogane:2016zbf,Shirogane:2020muc,Borsanyi:2022xml} and was also introduced for the discontinuity
of $\chi$ in Ref.~\cite{Lucini:2004yh}.

Though the splitting can be defined both for bare
and renormalized Polyakov loops, we prefer to work
with renormalized quantities. The details for
the renormalization procedure can be summarized as follows. The absolute value $|P|$ is defined as
\begin{equation}
    P(T;N_x,N_\tau) = P_0(\beta(T N_\tau);N_x,N_\tau) Z(\beta(T N_\tau))^{N_\tau}
    \label{eq:ploopZ}
\end{equation}
where $N_x$ and $N_\tau$ specify the lattice volume
and $P_0(\beta(T N_\tau); N_x, N_\tau)$ is the ensemble
average of the volume averaged bare Polyakov loop
at the given parameters. $\beta(T N_\tau)$ is the
gauge coupling at the $a^{-1} = T N_\tau$ scale.
The renormalization factor $Z(\beta)$ 
is determined by setting
a renormalization condition $P(T)\equiv1$ at $T=T_c$.
Thus, we can calculate $Z(\beta)$ at the $\beta_c(N_\tau)$ values. We determined $Z$ using $LT=4$ lattices with $N_\tau=5,6,7,8,10$ and 12. A polynomial 
fit to $\log Z(\beta)$ allows an interpolation in $\beta$. In the following systematic analysis the error coming from the Polyakov loop renormalization refers to the ambiguity in the $Z(\beta)$ interpolation scheme.

\begin{figure}
    \centering
    \includegraphics[width=\linewidth]{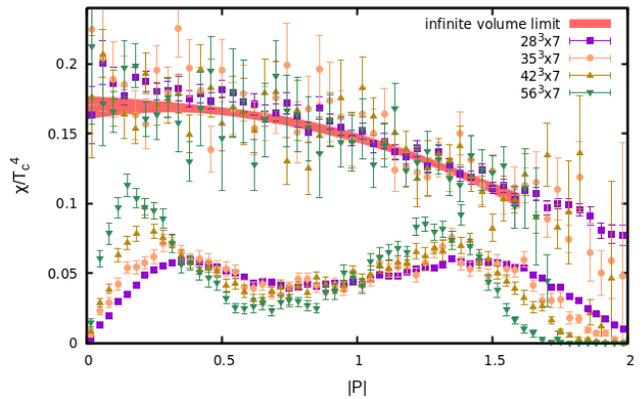}
    \caption{Topological susceptibility as a function of the absolute value of the renormalized Polyakov loop. The red curve is the infinite volume limit obtained from a two dimensional fit. In the lower region of the figure we show Polyakov loop histograms belonging to different lattice volumes. The temporal extension of the lattices used for this figure is $N_\tau=7$. The curves look similar for other lattices with $N_\tau=6,8,10$.}
    \label{fig:ploop-chi-hist}
\end{figure}

We illustrate the behaviour of the topological
fluctuations at $T_c$ in Fig.~\ref{fig:ploop-chi-hist}.
We show data for four volumes taken at $N_\tau=7$.
The lower curves are the histograms of $|P|$.
The cut value $P_c$ is the fitted local minimum 
between the peaks for the respective volume.

The data in Fig.~\ref{fig:ploop-chi-hist} and the complete
data set used in this and the next section are
taken using the tempering algorithm in a narrow range
around $\beta_c$. We stored only the configurations
simulated at $\beta_c$. Since $\beta_c$ itself has
an error, we reweighted our stored ensemble such
that the expectation value of the third Binder
cumulant of the bare Polyakov loop exactly vanishes.
In a jackknife-based error analysis this means that
for every jackknife sample a slightly
different $\beta_c$ was used.

The top curves in Fig.~\ref{fig:ploop-chi-hist} show the
topological susceptibility for each Polyakov loop bin
of the reweighted ensembles. We observe a smooth
function for each volume, with a mild volume dependence.
We extrapolated the infinite volume limit of this dependence with a 2D fit of a second degree polynomial. At $T=T_c$ the infinite volume Polyakov loop histogram is a double Dirac delta. Our results show that $\chi$ is a decreasing function of $|P|$. It also suggests that there should be a discontinuity at the transition temperature in the following way: In the infinite volume case we would have to subtract from  the value of the function $\chi(|P|)$ at the position of the ``deconfined peak'' the value in the confined phase $\chi(|P|=0$). In finite volumes, however, the Polyakov loop histogram does not have a sharp distinction between the two phases, therefore we have to define what we mean by one configuration being in one or the other phase. The Polyakov loop histograms have two peaks that become sharper as we increase the volume. A natural way to identify which phase the configurations belong to is to cut $P_c$ the Polyakov loop histogram at its minimum between the two peaks.
Then we can assign topological susceptibilities to both phases for each ensemble.

\begin{figure}
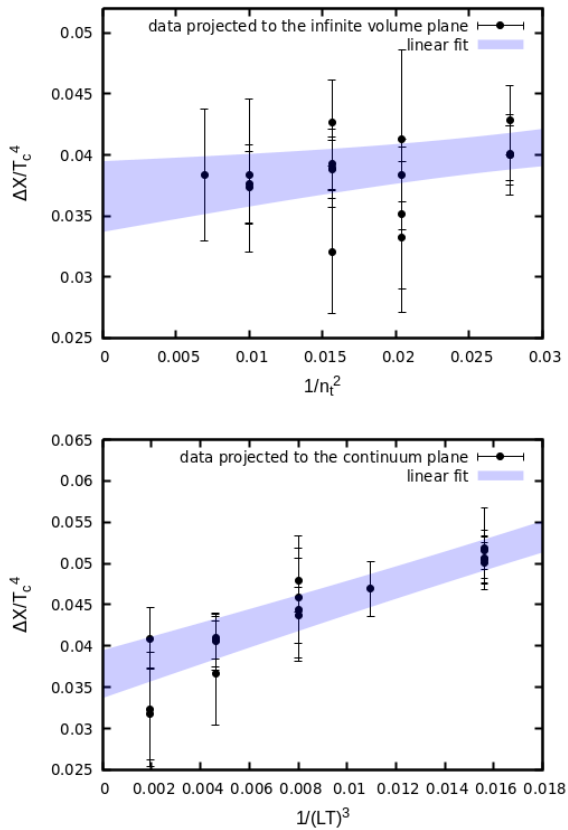

    \centering
    \includegraphics[width=1.1\linewidth]{infV_proj_DChi.pdf}
    \includegraphics[width=1.1\linewidth]{cont_proj_DChi.pdf}
    \caption{Discontinuity of $\chi$ at the transition temperature of ensembles listed in Table \ref{tab:1}. Results of different lattices are projected onto the infinite volume plane (top) and the continuum plane (bottom). The blue bands are linear (three parameter) fits of the projected data using the same parameters as in the two dimensional fit.}
    \label{fig:ploop-chi}
\end{figure}

We determined $\Delta \chi$ at the transition ensemble-by-ensemble by subtracting the value of $\chi$ in cold phase from that of the hot phase. Then we extrapolated the infinite volume and the continuum limit via a two dimensional fit. In Fig. \ref{fig:ploop-chi} we show a linear fit on data projected to the infinite volume plane (top panel) and data projected to the continuum plane (bottom panel). 
The main result for the discontinuity of the topological susceptibility with the statistical and systematic errors is shown in table \ref{tab:delta_chi}. The systematic error is coming from the following four systematic variables. First we varied the fit range by including and excluding data with the smallest aspect ratio $LT=4$, then we used two different fit formulas for the infinite volume and continuum extrapolations, one was a function with three parameters $f(x,y)=a+b\cdot x+c\cdot y$ and the other was a function with four parameters $g(x,y)=a+b\cdot x+c\cdot y+d\cdot xy$, with $x=1/N_\tau^2$ and $y=1/(LT)^3$. Furthermore we varied the fit range of the function that was used to determine the minima of Polyakov loop histograms, the smaller range being $0.15|P|-1.85|P|$ and the larger range was $0.1|P|-1.9|P|$. Finally we used two different schemes to interpolate the renormalization factors of the Polyakov loop. As expected, the systematics is dominated by the ambiguities in the infinite volume extrapolation.

Our directly calculated result $\Delta\chi/T_c^4 = -0.0344(44)(32)$ agrees with the estimated discontinuity of $\chi$ obtained from equation \ref{eq:clapeyron}.

\begin{table}[]
\centering
\begin{tabular}{l|l|l}
\hline\hline
\multicolumn{3}{c}{$\Delta\chi/T_c^4$} \\ \hline
median                         & \multicolumn{2}{c}{-0.034378} \\ \cline{2-3}
statistical error              &   0.0044 &   13 \% \\
full systematic error          &   0.0032 &  9.3 \% \\
\hline
Fit range                      &    0.0026 &   7.43 \% \\
Fit formula                    &    0.0026 &   7.54 \% \\
Fit range of histogram         &    0.0000 &   0.05 \% \\
Renormalizing                  &    0.0000 &   0.11 \% \\
\hline\hline
\end{tabular}
\caption{\label{tab:delta_chi}Result for the discontinuity of the topological susceptibility with its statistical (second row) and systematic (third row) errors. In the bottom four rows we show constituents of the systematic error. From top to bottom these are errors coming from the change in results by including data with $LT=4$ or not, the use of different fit formulas for the two dimensional extrapolation, the range of the fit on Polyakov-loop histograms when calculating their minima and the change in the renormalizing factor $Z$ of $|P|$.}
\end{table}


\section{\label{sec:curvature}The $\theta$-dependence of the transition temperature}

As we mentioned in the introduction, and was explained in the study of the Pisa group of Ref.~\cite{DElia:2012pvq} (see Eqs.~(\ref{eq:Rtheta}-\ref{eq:clapeyron})), the discontinuity of the topological susceptibility at $T_c$ is linked, through the latent heat, to the curvature of the first order line in the $\theta{-}T$ phase diagram.

The method to determine $R_\theta$ in Ref.~\cite{DElia:2012pvq} uses simulations at imaginary values of the $\theta$ parameter ($\theta^I$). This procedure is very similar to the study of $T_c$ as a function of the chemical potential in full QCD, where, again, the use of imaginary chemical potentials is one of the standard techniques \cite{Bonati:2015bha,Bellwied:2015rza,Cea:2015cya,Borsanyi:2020fev}.

Just like with $\theta$, the curvature can alternatively be obtained
employing high statistics $\mu_B=0$ ensembles
\cite{Bazavov:2018mes}, and the equivalence of the two approaches can be demonstrated \cite{Bonati:2018nut}.

In most of these works the transition temperature was identified as the peak of a susceptibility
(Polyakov loop in the SU(3) theory, and the chiral susceptibility in full QCD). In the case of the SU(3) Yang-Mills theory we have exploited in Ref.~\cite{Borsanyi:2022xml} the definition of the transition temperature as the location where $b_3(\beta_c)=0$, $b_3$ being the third Binder cumulant of the Polyakov loop absolute value.
This third order cumulant could be obtained with high precision thanks to parallel tempering. The zero-crossing of $b_3$ was found through reweighting in $\beta$ from a single gauge coupling, since all streams in the tempered simulation are roughly equally represented at each $\beta$. This eliminated the need for fitting the curve $b_3(\beta)$.

Analogously to the study of the $T{-}\mu_B$ phase diagram, we can also extract $R_\theta$ from $\theta=0$
ensembles. To achieve this we start from the
sub-ensemble of the tempered $\theta=0$ simulation corresponding to $\beta\approx\beta_c(0)$, that we already used to
obtain $\Delta\chi$. In these sub-ensemble $Q$ was determined using the gradient flow.
We perform a simultaneous reweighting in $\theta$
and $\beta$ in order to maintain $b_3=0$.
The ratio $\Delta\beta/\Delta\theta$ can then be used to extract $R_\theta$ in the $\Delta\theta\to0$ limit (in practice, we used a very small value of $\theta= 0.02i$).

In addition to this, we performed simulations at imaginary $\theta$.
In previous works the Hybrid-Monte-Carlo algorithm has been often used  to sample $Q$-dependent actions, where for $Q$ a proxy charge was introduced
\cite{DElia:2012pvq,Bonati:2018blm,Borsanyi:2021gqg}.
The proxy charge was often a non-smeared clover expression or one with lesser smearing. The difference between the proxy and the actually used charge definition was taken into account through multiplicative renormalization \cite{Panagopoulos:2011rb}.

Instead of the Hybrid Monte-Carlo technique we use the pseudo-heatbath algorithm (with overrelaxation sweeps) to propose updates that undergo a Metropolis step to accept or reject the update according to the action $S_\textrm{topo} = Q\theta^I$. This $Q$ is defined using a sequence of stout smearings and the improved clover definition as described in Section~\ref{sec:stopo} and Appendix \ref{sec:data}, such that the renormalization step is no longer necessary. For modest $\theta^I$ parameters (e.g. $\theta^I<2$) and volumes $(LT\le6)$ we find reasonable acceptance ($>$10\%). The range of accessible $\theta^I$ parameters diminishes with the inverse volume. This, however, does not prohibit the use of larger lattices, since the slope of $b_3(\beta)$ scales proportionally to the volume, increasing the achievable precision on $\beta_c(\theta^I)$ accordingly.
Thus, in a larger volume we can extract $R_\theta$ with a smaller lever arm (a smaller value of $\theta^I$).
Considering that the Hybrid-Monte-Carlo algorithm is at least 10$\times$ less efficient for the Yang-Mills theory than the heatbath update, even without counting the costs for the $Q$-dependent forces, we see this strategy as a resource-saving alternative. 

The following proxy quantity can be defined both at $\theta=0$ an well as for imaginary $\theta$\,:

\begin{equation}
\frac{\frac{T_c(\theta)}{T_c(0)} - 1}{\theta^2}=
 \mathcal{F}\left(\theta^2, 1/N_\tau^2,1/(LT)^3\right)
\end{equation}

Its value at vanishing arguments is $R_\theta$ in the thermodynamic and continuum limits.

In total we used 40 ensembles (see Table~\ref{tab:imtheta} in appendix \ref{sec:data}). We perform a global fit to the data $\mathcal{F}(x,y,z)= R_\theta+Ax+By+Cz$, where $A$, $B$ and $C$ are the leading slopes for the residual $\theta^2$, lattice spacing and volume dependence of $\mathcal{F}$, respectively.

We consider three sources of systematic errors.
First, the scale setting ambiguity, here using different interpolations to the $w_0a(\beta)$ function.
Second, we varied the fit formula, by enabling or disabling the $C/(LT)^3$ term. Most importantly, the third option controlled the continuum limit range: whether we included or excluded the coarsest lattice $N_\tau=6$ in the continuum extrapolation.
Finally we arrive at:

\begin{center}
\begin{tabular}{l|l|l}
\hline\hline
\multicolumn{3}{c}{$R_\theta$}\\
\hline
median                         & \multicolumn{2}{c}{0.0181} \\
statistical error              &  0.00045 &  2.5 \% \\
full systematic error          &  0.00064 &  3.5 \% \\
\hline
w0 interpolation               &    $5\cdot10^{-6}$ &   0.03 \% \\
choice of the fit function $\mathcal{F}$      &    0.00003 &   0.14 \% \\
continuum extrap. range        &    0.0006 &   3.5 \% \\
\hline\hline
\end{tabular}
\end{center}

This result is in remarkable agreement with the earlier continuum extrapolated (though not infinite volume extrapolated) value given by the Pisa group 0.0178(5) \cite{DElia:2013eua}.


\section{\label{sec:b2}On the kurtosis of the topological charge distribution}

Just above $T_c$, the structure of the topological fluctuations of pure ${\rm SU}(3)$ undergoes a significant transition from a dense medium without discrete localizations of charge to an ideal gas of sparse lumps of charge described by the dilute instanton gas approximation (DIGA) \cite{Vig:2021oyt}. How close ``just above'' is, however, has been an uncertain matter, but recent work has shown that the structure of topological objects is already consistent with that of an ideal gas between $1.045 T_c$ and $1.15 T_c$ \cite{Vig:2021oyt, Bonati:2013tt}.

The DIGA model makes distinct predictions for the values of high-order cumulants of the topological charge such as the kurtosis $b_2$, which is then a useful quantity in the determination of the onset of the ideal gas behavior. In the infinite volume limit, the topological charge in the DIGA model follows a Skellam distribution, \cite{Vig:2021oyt}, \begin{equation}
    P(Q) = e^{-(\mu_i+\mu_a)} I_Q(2\sqrt{\mu_i \mu_a}) = e^{-V\chi} I_Q(V\chi), \label{eq:skellam}
\end{equation} where $\mu_i$ and $\mu_a$ are the means of the independent Poisson distributions of instantons and anti-instantons respectively, and $\mu_i = \mu_a = \mathcal{V}\chi/2 = \left\langle Q^2 \right\rangle/2$. From Eq.~(\ref{eq:b2}), $b_2$ has the analytic value of $-1/12$ for this distribution. At $T = 0$, empirical results on the lattice indicate that $b_2$ assumes a value of approximately $-0.02$ \cite{Bonati:2015sqt, Panagopoulos:2011rb, Ce:2015qha, DElia:2003zne, Giusti:2007tu}, and does not depart much from this value for $T < T_c$. How quickly the onset of the DIGA picture occurs across the phase transition can then be seen in how $b_2(T)$ departs from this empirical value and approaches the DIGA limit.

Unfortunately, the determination of $b_2(T)$ is hampered by the requirement of very high statistics for the precise measurement of the fourth moment of the topological charge, $\left\langle Q^{4} \right\rangle$, which is what makes direct measurement of the kurtosis difficult at large volumes. We saw in Fig.~\ref{fig:chiANDb2_LT4_cont} how the errors for $b_2$ are much larger than for $\chi$. Finite-volume effects obscure whether $b_2$ approaches the DIGA limit from above or below.

To peel away some of the uncertainty in the behavior of $b_2$ due to the low statistics, we derive an identity for $b_2$ and apply a model that allows us to reconstruct the $b_2$ using the more easily measurable topological susceptibility $\chi$. (The derivation of this identity is shown in detail in Appendix \ref{sec:kurtosis identity}.) At a given $T$, $b_2$ can be written as 
\begin{align}
    b_2  &= \frac{\int dP \, b_2(P) \chi(P) \rho(P)}{\int dP \, \chi(P) \rho(P)}  \nonumber \\
    &\quad - \frac{\mathcal{V}}{4} \frac{\int dP \, \chi^2(P) \rho(P) - \left(\int dP \, \chi(P) \rho(P)\right)^2}{\int dP \, \chi(P) \rho(P)} \, \, , \label{eq:b2identity}
\end{align} 
where $P$ is the absolute value of the renormalized Polyakov loop, $\rho(P)$ is the distribution of $P$ values in the ensemble (in practice, a histogram), while $\chi(P)$ and $b_2(P)$ are respectively the susceptibility and kurtosis as functions of $P$ at fixed temperature. In an ensemble at a given temperature, by binning the charge $Q$ according to the values of $P$, $\chi(P)$ and $b_2(P)$ can be easily determined bin by bin. However, the issue of low statistics becomes exacerbated for $b_2(P)$ due to the binning, making it unfeasible to compute $b_2(P)$ directly.

\begin{figure}
    \centering
    \includegraphics[trim=10 10 20 10, clip, width=\linewidth]{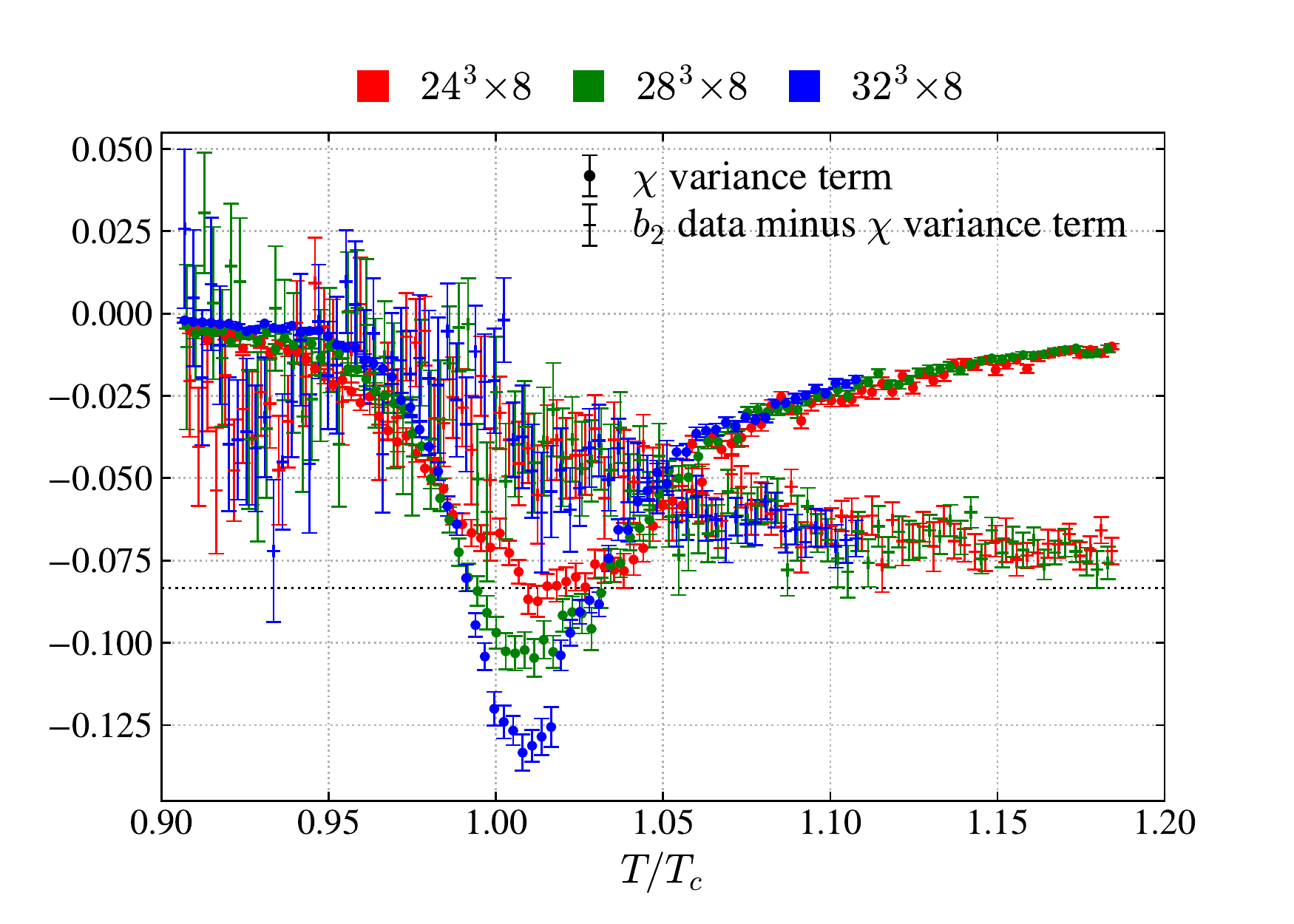}
    \caption{The second term from Eq.~(\ref{eq:b2identity}) containing the variance of $\chi(P)$ for three $N_\tau = 8$ lattices as a function of temperature, as well as this term subtracted from the $b_2(T)$ data. The $\chi(P)$ term contains the volume dependence of $b_2(T)$.}
    \label{fig:b2 volume dependence}
\end{figure}

In Fig. \ref{fig:b2 volume dependence}, we plot the two terms of Eq. (\ref{eq:b2identity}) separately for three lattice volumes at $N_\tau = 8$. The first term, containing $b_2(P)$, is found by subtracting the second term, containing the variance of $\chi(P)$, from the $b_2(T)$ data.  The crucial point is that, while the second term shows clear volume scaling, the first one does not, which indicates that the volume dependence of the kurtosis is isolated within the $\chi(P)$ term. If we expect a discontinuity in $\chi(T)$ at $T = T_c$ in the thermodynamic limit, the $\chi(P)$ term becomes a downward delta function at $T = T_c$ in this limit; the $b_2(P)$ term in Fig.~\ref{fig:b2 volume dependence} may, therefore, be the thermodynamic limit of $b_2(T)$. In that case, the kurtosis approaches the DIGA limit gradually from above across the transition. We compared the $b_2(P)$ terms of other lattices at $N_\tau = 6$ and $7$ and found that they lie on top of the same curve as the $N_\tau = 8$ lattices.

Assuming, then, that the volume dependence of $b_2(P)$ is negligible, as well as the temperature and cutoff dependence, we substituted for $b_2(P)$ a simple rational ansatz: \begin{equation}
    b_2(P) = -\frac{a_0 + a_1 P + a_2 P^2}{1 + c_1 P + c_2 P^2}. \label{eq:b2ansatz}
\end{equation} We treated this as a lowest-order approximation of the true $b_2(P)$.  We enforced the constraint that $c_2 = 12a_2$, motivated by the following thought. For $T > T_c$, as $b_2$ approaches the DIGA limit of $-1/12$, the distribution of $P$ values, $\rho(P)$, becomes a single peak located at $P > 1$. The second term of Eq. (\ref{eq:b2identity}), containing the variance of $\chi(P)$, only significantly contributes at $T \approx T_c$ when $\rho(P)$ shows two peaks, and vanishes otherwise leaving only the $b_2(P)$ term. Thus, since $T > T_c$ corresponds to sampling increasingly from $P > 1$, in order for $b_{2,{\rm model}} \rightarrow -1/12$ at large $T$, $\lim_{P\rightarrow\infty} b_2(P) = -1/12$.

\begin{figure*}
    \centering
    \includegraphics[trim=10 10 50 10, clip, width=\linewidth]{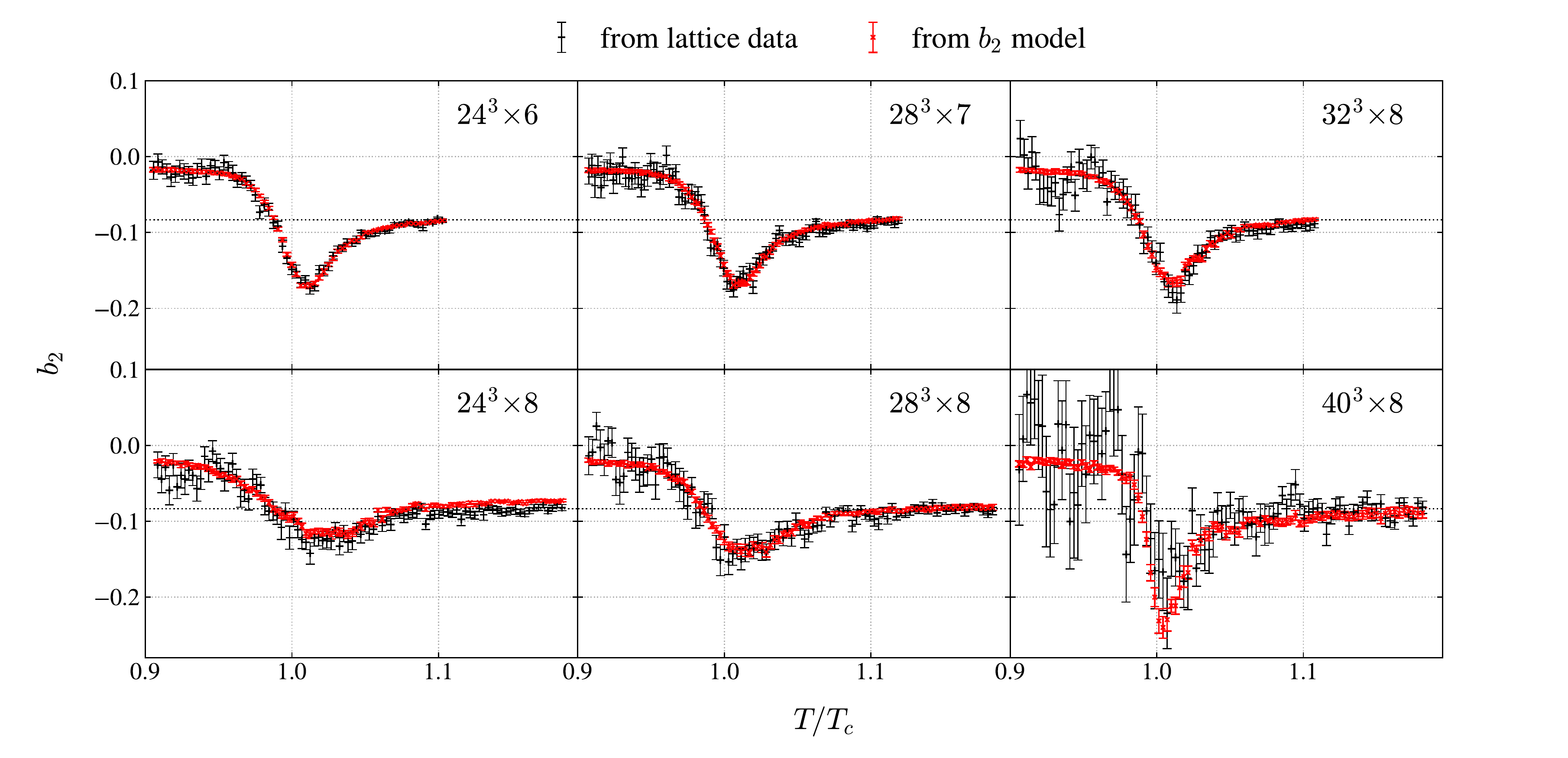}
    \caption{The kurtosis $b_2$ of the topological charge distribution as a function of the normalized temperature for several lattices computed (black) directly from the topological charge data and (red) using the identity in 
    Eq.~(\ref{eq:b2identity}) with a rational ansatz for $b_2(P)$. The parameters of the model were found by fitting to the $b_2$ data from the $24^3{\times}6$ lattice. These parameters were used to compute $b_2$ using the model for all the other lattices. The DIGA limit $b_2 = -1/12$ is indicated with a dashed line.}
    \label{fig:b2 model examples}
\end{figure*}

We fitted the parameters of $b_2(P)$ for a particular lattice by minimizing \begin{equation}
    R^2 = \sum_{T} \frac{(b_{2,{\rm data}}(T) - b_{2}(T))^2}{\sigma^2(T)}, \label{eq:minimize}
\end{equation} where the sum is over all the simulated $T$ values and $\sigma$ is the jackknife error of the $b_2(T)$ data. Because the $24^3{\times}6$ lattice had the most statistics, we used its fit parameters to reconstruct $b_2(T)$ for several lattices using Eq. (\ref{eq:b2identity}), which is shown in Fig. \ref{fig:b2 model examples}.  As a first approximation, neglecting temperature, volume, and cutoff effects on $b_2(P)$, the reconstructed $b_2(T)$ follows the shape of the data quite well and helps resolve the volume dependence of the kurtosis more clearly at large volumes. Indeed, the volume dependence is the surest part of the reconstructed $b_2(T)$, since the second term of 
Eq.~(\ref{eq:b2identity}) is analytic and independent of the ansatz.

\begin{figure}
    \centering
    \includegraphics[trim=5 5 30 40, clip, width=0.8\linewidth]{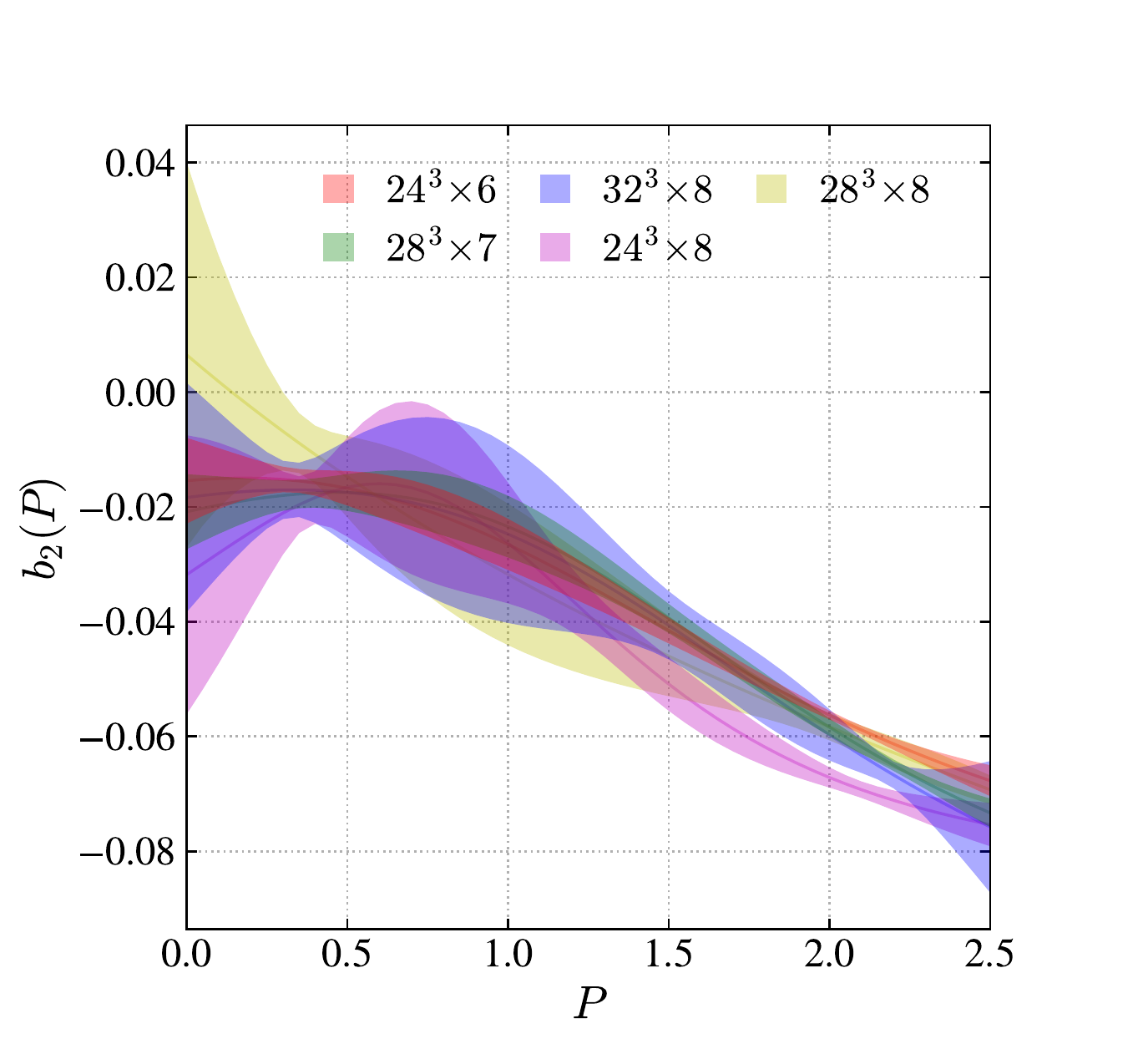}
    \caption{The kurtosis $b_2$ as a function of the absolute value of the (renormalized) Polyakov loop fitted to the $b_2(T)$ data for several lattices by minimizing Eq. (\ref{eq:minimize}).}
    \label{fig:b2 vs P}
\end{figure}

To further demonstrate that $b_2(P)$ is largely independent of volume and the lattice spacing, we fitted the parameters of $b_2(P)$ using the $b_2(T)$ data of other lattices with good statistics. Fig. \ref{fig:b2 vs P} shows the results of these fits.  The $b_2(P)$ curves from all five lattices lie in general agreement with one another. The parameter $a_0$ tended not to be constrained very well due to the low statistical weight at $P = 0$, resulting in the ``horn'' shape of the plot in Fig. \ref{fig:b2 vs P}; however, it is worth noting that for $24^3{\times}6$, the fit was constrained enough to yield $a_0 = 0.0155(75)$, which is in surprising agreement with the empirical results from Ref. \cite{Bonati:2015sqt, Panagopoulos:2011rb, Ce:2015qha, DElia:2003zne, Giusti:2007tu}. The pinch in the plot at the base of the horn is where the fit was heavily constrained by the peak of $\rho(P)$ corresponding to the confined phase.


\section{\label{sec:conclusions}Conclusions}

In this work we studied the distribution of the topological charge in the SU(3) Yang-Mills theory within the framework of lattice QCD. The first order transition manifests itself in the discontinuity of several observables, most notably, the Polyakov loop and the energy density, but also the topological susceptibility.

Our investigations are centered around the three
quantities linked by Eq.~(\ref{eq:clapeyron}) \cite{DElia:2012pvq}.

We have performed simulations of the Symanzik improved gauge action in the vicinity of the phase transition
temperature, using parallel tempering to reduce autocorrelations. 
For the topological density the Symanzik improved clover definition was used. To negate cutoff effects,
we have defined the physical topological charge $Q$ at a finite physical Wilson flow time to allow for continuum extrapolations. A simplified definition of $Q$ using stout smearing steps to approximate the Wilson flow was also
used, after confirming that this choice gives rise to negligible systematic errors.

Similarly to the latent heat \cite{Borsanyi:2022xml}, the drop of the topological susceptibility across the deconfinement 
transition can be measured by noticing 
that the average susceptibility has a smooth dependence on the average Polyakov loop variable with mild volume dependence. The 
discontinuity may then be read off as the values of the average susceptibility at the peaks of the 
Polyakov loop distribution. Alternatively, one separates all configurations into the confined and deconfined phase by a cut in the Polyakov loop, corresponding to the minimum of the double-peaked histogram.
In the thermodynamic limit and at the transition temperature this distribution is a double Dirac delta, corresponding to the two distinct phases of the theory.
Only in this limit can we define $\Delta\chi$ unambiguously. For the volume extrapolation we use lattices with an aspect ratio $LT=N_x/N_\tau$ up to 8.
Our continuum and infinite volume extrapolated result is $ \Delta \chi / T_c^4 =- 0.0344(44)(32)$.

We have also measured the $R_\theta$ parameter by investigating the phase transition at finite imaginary $\theta$ parameters by performing reweighting of simulations at $\theta=0$, as well as simulations at 
$ \textrm{Im} \theta>0$ taking the topological term into account using an extra accept-reject step. 
Our continuum and infinite volume extrapolated result is $ R_\theta= -0.01810(45)(64) $.

The behavior of the $b_2$ parameter of the topological charge distribution was also studied across the 
phase transition. At low temperatures $b_2$ is roughly constant with the value $\approx -0.02$ \cite{Bonati:2015sqt}. At high temperatures
it converges to $ b_2=-1/12$, which can be understood in terms of the DIGA (dilute instanton gas approximation) picture. Using an identity for $b_2$ in terms of binned averages as a function of the 
Polyakov average, we have successfully identified the main source of the volume dependence in $b_2(T)$, 
suggesting that in the infinite volume limit, $b_2$ approaches the DIGA limit from above. On realistic system volumes, however, we observe the dominance of a negative delta peak, which is due to phase coexistence near $T_c$.

Finally, using our result $\Delta\epsilon/T_c^4 = 1.025(21)(27)$ for the latent heat from a recent 
study \cite{Borsanyi:2022xml}, we can confirm the validity of the relation 
$ 2 \Delta \epsilon R_\theta  = \Delta \chi  $, which in this form can be used for the estimation 
of the discontinuity of the susceptibility to yield $ 2 \Delta \epsilon R_\theta /T_c^4= \Delta \chi/T_c^4 = -0.0371 $ with 11\% combined statistical and systematic errors.
The direct calculation yields the compatible result $ \Delta \chi / T_c^4 = - 0.0344 $, with 22\% 
combined statistical and systematic errors.

\textbf{Acknowledgements}

The project reveived support from the DFG under Grant. No. 496127839. 
This work is also supported by the MKW NRW under the funding code NW21-024-A.  
The authors gratefully acknowledge the Gauss Centre for Supercomputing e.V. (www.gauss-centre.eu) for funding this project by providing
computing time on the GCS Supercomputer HAWK at HLRS, Stuttgart and on the Juwels/Booster at FZ-Juelich. Part of the computation was performed on the cluster at the University of Graz.

\appendix

\section{\label{sec:kurtosis identity}Kurtosis identity}

The $n$th moment of the topological charge $Q$ at a fixed temperature $T$ on the lattice can be calculated as a weighted average via \begin{equation}
    \left\langle Q^{n} \right\rangle = \int dP \, \left\langle Q^{n} \right\rangle_P \rho(P), \label{eq:nth moment}
\end{equation} where $P$ is the Polyakov loop magnitude, $\rho(P)$ is the probability distribution function of the $P$ values of the lattice configurations computed in an ensemble at a given temperature, and $\left\langle Q^{n} \right\rangle_P$ is the $n$th moment of $Q$ among just those configurations with a Polyakov loop magnitude of $P$. We define the topological susceptibility and kurtosis as functions of $P$ at fixed temperature: 
\begin{eqnarray}
    \chi(P)   &= \frac{\left\langle Q^2 \right\rangle_P}{\mathcal{V}}, \label{eq:chi(P)}
\end{eqnarray} \begin{equation}
    b_2(P)  = - \frac{\chi_4(P)}{12 \chi(P)}, \quad \chi_4(P) = \frac{1}{\mathcal{V}} \left[ \left\langle Q^4 \right\rangle_P - 3 \left\langle Q^2 \right\rangle_P^2 \right]. \label{eq:b2(P)}
\end{equation} We recover the susceptibility in Eq. (\ref{eqn:chi}) directly via \begin{eqnarray}
    \chi &= \int dP \, \chi(P) \rho(P),
\end{eqnarray} and we recover $b_2$ in Eq. (\ref{eq:b2}), a nonlinear combination of moments, via \begin{align}
    b_2 &= - \frac{\int dP \, \left\langle Q^{4} \right\rangle_P \rho(P) - 3 \left(\int dP \, \left\langle Q^{2} \right\rangle_P \rho(P)\right)^2}{12\mathcal{V} \int dP \, \chi(P) \rho(P)} \nonumber \\
    &= - \frac{\int dP \, \left\langle Q^{4} \right\rangle_P \rho(P) - 3\mathcal{V}^2 \left(\int dP \, \chi(P) \rho(P)\right)^2}{12\mathcal{V} \int dP \, \chi(P) \rho(P)}. 
    \label{eq:b2recover}
\end{align} $\left\langle Q^{4} \right\rangle_P$ can be eliminated by solving Eq. (\ref{eq:b2(P)}) for $\left\langle Q^{4} \right\rangle_P$ and then inserting this into 
Eq.~(\ref{eq:b2recover}) to yield \begin{align}
    b_2  &= \frac{\int dP \, b_2(P) \chi(P) \rho(P)}{\int dP \, \chi(P) \rho(P)} \nonumber \\
    &\quad - \frac{\mathcal{V}}{4} \frac{\int dP \, \chi^2(P) \rho(P) - \left(\int dP \, \chi(P) \rho(P)\right)^2}{\int dP \, \chi(P) \rho(P)}.
\end{align} 

\section{\label{sec:data} Tabulated data}

In this Appendix we give some of the intermediate simulation results that entered our analyses. Table~\ref{tab:deltachidata} was used in Section~\ref{sec:jump} and Table~\ref{tab:imtheta}
entered the fits in Section~\ref{sec:curvature}.

\begin{table}[h]
\begin{center}
\begin{tabular}{||c|c||}
\hline
lattice & $\Delta\chi/T_c^4$ \\
\hline
$24^3\times6$ & 0.0548(7)\\
$28^3\times7$ & 0.0529(18)\\
$32^3\times8$ & 0.0534(16)\\
$40^3\times10$ & 0.0522(23)\\
$48^3\times12$ & 0.0526(60)\\
$36^3\times8$ & 0.0488(27)\\
$30^3\times6$ & 0.0473(32)\\
$35^3\times7$ & 0.0484(65)\\
$40^3\times8$ & 0.0497(25)\\
$50^3\times10$ & 0.0457(52)\\
$36^3\times6$ & 0.0450(19)\\
$42^3\times7$ & 0.0390(56)\\
$48^3\times8$ & 0.0429(20)\\
$60^3\times10$ & 0.0413(27)\\
$48^3\times6$ & 0.0440(27)\\
$56^3\times7$ & 0.0347(60)\\
$64^3\times8$ & 0.0351(47)\\
\hline
\end{tabular}
\end{center}
\caption{\label{tab:deltachidata}
The discontinuity in the topological susceptibility ($\Delta\chi/T_c^4$) on our lattices that entered the combined infinte volume and continuum limit.
$\Delta\chi/T_c^4$ is defined as the topological susceptibility difference calculated from configurations above and below a Polyakov loop cut.
}
\end{table}

\begin{table}[h]
\begin{center}
\begin{tabular}{||cc||cc||}
\hline
$\theta^I$ & $\beta_c$ & $\theta^I$ & $\beta_c$\\
\hline
\multicolumn{4}{||c||}{$24^3\times6$}\\\hline
0.50 & 4.31472(19) &
0.75 & 4.31827(19)\\
1.00 & 4.32300(19) &
1.25 & 4.32935(24)\\
1.50 & 4.33668(26) & 
\multicolumn{2}{c||}{}\\
\hline
\multicolumn{4}{||c||}{$28^3\times7$}\\\hline
0.75 & 4.42107(27) &
1.00 & 4.42640(20)\\
\hline
\multicolumn{4}{||c||}{$30^3\times6$}\\\hline
0.50 & 4.31598(19) &
0.75 & 4.31965(17)\\
1.00 & 4.32463(22) &
\multicolumn{2}{c||}{}\\
\hline\multicolumn{4}{||c||}{$32^3\times8$}\\\hline
1.00 & 4.51965(26) &
1.25 & 4.52629(40)\\
1.50 & 4.53509(40) &
2.00 & 4.55513(40)\\
\hline\multicolumn{4}{||c||}{$36^3\times6$}\\\hline
0.50 & 4.31611(24) &
0.75 & 4.31949(25)\\
1.00 & 4.32446(37) &
\multicolumn{2}{c||}{}\\
\hline
\multicolumn{4}{||c||}{$36^3\times8$}\\\hline
1.00 & 4.52083(22) &
1.20 & 4.52570(50)\\
\hline
\multicolumn{4}{||c||}{$40^3\times10$}\\\hline
0.50 & 4.67312(77) &
0.75 & 4.67872(55)\\
1.00 & 4.68307(52) &
\multicolumn{2}{c||}{}\\
\hline
\multicolumn{4}{||c||}{$40^3\times8$}\\\hline
1.00 & 4.52096(40) &
1.25 & 4.52858(45)\\
1.50 & 4.53618(37) &
\multicolumn{2}{c||}{}\\
\hline
\multicolumn{4}{||c||}{$48^3\times6$}\\\hline
0.40 & 4.31654(24) &
0.50 & 4.31618(18)\\
0.75 & 4.31351(30) &
\multicolumn{2}{c||}{}\\
\hline
\multicolumn{4}{||c||}{$48^3\times8$}\\\hline
0.75 & 4.51681(17) &
1.25 & 4.52923(22)\\
\hline
\multicolumn{4}{||c||}{$60^3\times10$}\\\hline
0.75 & 4.68067(25) &
1.00 & 4.68613(27)\\
\hline
\end{tabular}
\end{center}
\caption{\label{tab:imtheta}Gauge couplings at the transition temperature for various lattices and imaginary $\theta$ parameters.}
\end{table}

\twocolumngrid

\bibliography{thermo}

\end{document}